\documentclass{article}

\usepackage{arxiv}
\usepackage{orcidlink}      % fors ORCID
\usepackage[utf8]{inputenc} % allow utf-8 input
\usepackage[T1]{fontenc}    % use 8-bit T1 fonts
\usepackage{hyperref}       % hyperlinks
\usepackage{url}            % simple URL typesetting
\usepackage{booktabs}       % professional-quality tables
\usepackage{amsfonts}       % blackboard math symbols
\usepackage{nicefrac}       % compact symbols for 1/2, etc.
\usepackage{microtype}      % microtypography
\usepackage{graphicx}
\usepackage{algpseudocode}
\usepackage{amssymb}
\usepackage{natbib}
\usepackage{authblk}
\usepackage[colorinlistoftodos]{todonotes}
\usepackage{doi}
\usepackage{authblk}
\usepackage{fdsymbol}
\usepackage{caption}
\newenvironment{acknowledgements}{\section*{Acknowledgements}}{}

\renewcommand{\algorithmicforall}{\textbf{for each}}
\newcommand{\var}{\texttt}
\newcommand{\func}{\mathrm}

\usepackage{algorithm}
\begin{document}

\title{
MAIR: Framework for mining relationships between research articles, strategies, and regulations in the field of explainable artificial intelligence}

\author[2,1]{\orcidlink{0000-0002-3695-9809}Stanisław Giziński}
\author[2,1]{\orcidlink{0000-0002-9181-0126}Michał Kuźba}
\author[3]{\orcidlink{0000-0003-2664-2135}Bartosz Pieliński}
\author[4]{\\\orcidlink{0000-0003-2097-1499}Julian Sienkiewicz}
\author[5]{\orcidlink{0000-0003-0563-7855}Stanisław Łaniewski}
\author[1,2]{\orcidlink{0000-0001-8423-1823}Przemysław Biecek}
\affil[1]{ MI$^2$ Data Lab, Faculty of Mathematics and Information Science, Warsaw University of Technology}
\affil[2]{ Faculty of Mathematics, Informatics and Mechanics, University of Warsaw}
\affil[3]{ Faculty of Political Science and International Studies, University of Warsaw}
\affil[4]{ Faculty of Physics, Warsaw University of Technology }
\affil[5]{ Quantitative Psychology and Economics, Faculty of Economics, University of Warsaw}
\maketitle

\captionsetup[figure]{labelfont={bf},name={Fig.},labelsep=period, font=small}
\captionsetup[table]{labelfont={bf},name={Table},labelsep=period, font=small}

\renewcommand{\algorithmicforall}{\textbf{for each}}

\begin{abstract}
The growing number of AI applications, also for high-stake decisions, increases the interest in Explainable and Interpretable Machine Learning (XI-ML). This trend can be seen both in the increasing number of regulations and strategies for developing trustworthy AI and the growing number of scientific papers dedicated to this topic. To ensure the sustainable development of AI, it is essential to understand the dynamics of the impact of regulation on research papers as well as the impact of scientific discourse on AI-related policies. 
This paper introduces a novel framework for joint analysis of AI-related policy documents and eXplainable Artificial Intelligence (XAI) research papers. The collected documents are enriched with metadata and interconnections, using various NLP methods combined with a methodology inspired by Institutional Grammar. Based on the information extracted from collected documents, we showcase a series of analyses that help understand interactions, similarities, and differences between documents at different stages of institutionalization.
To the best of our knowledge, this is the first work to use automatic language analysis tools to understand the dynamics between XI-ML methods and regulations. We believe that such a system contributes to better cooperation between XAI researchers and AI policymakers.

% Artificial intelligence methods are playing an increasingly important role in global economics. The growing importance and, at the same time, the risks associated with AI are driving a vibrant discussion about the responsible development of artificial intelligence. This results in both new computational methods as well as in new strategies and regulations. 

% However, the analysis of these documents shows a gap between the expectations (enshrined in strategies and regulations) and the reality (presented in research papers) related to the perception of AI. Due to the increasing number of papers, it is impossible to analyse this gap by manually analysing all the source documents.

% In this paper, we propose a new algorithm to automatically analyse a growing number of documents against the perception of AI. This algorithm is based on Institutional Grammar enriched with state-of-the-art techniques from the field of Natural Language Processing (NLP). 

% We apply the proposed algorithm to a set of XXX articles on AI extracted from arxiv and a set of YYY strategies extracted from ...
% The final contribution of this work is a discussion of the differences in perceptions of AI between different types of documents, different countries, and different sectors of the economy.

% To the best of our knowledge, this is the first work to use automatic language analysis tools to analyse regulation and  perception of AI.
\end{abstract}

% keywords can be removed
\keywords{Institutional Grammar \and Responsible Artificial Intelligence \and Monitoring regulations \and Knowledge Extraction}

\section{Introduction}
Artificial intelligence methods are playing an increasingly important role in global economics. The growing importance and, at the same time, the risks associated with AI are driving a vibrant discussion about the responsible development of artificial intelligence. Examples of negative consequences resulting from black-box models show that interpretability, transparency, safety, and fairness are essential yet sometimes overlooked components of AI systems.
Efforts to secure the responsible development of AI systems are ongoing at many levels and in many communities, both policymakers and academics \citep{gill_responsible_2020, barredo_arrieta_explainable_2020, baniecki_dalex_2020}.

Naturally, national strategies for the development of responsible AI, sector regulations related to the safe use of AI, as well as academic research related to new methods that ensure the transparency and verifiability of models are all interrelated. Strategies are based on discussions in the scientific community and are often sources of inspiration for subsequent research work. The need for regulation stems from risks, often identified by the research community, but when regulations are created, they become a powerful tool for developing methods to meet expectations. Scientific work in AI is particularly strongly connected to the economy, which means that a large part of it responds to the threads identified in regulations and strategies.

Although this impact is strong, we know little about the dynamics and structure of this impact. Analyses of AI-related policies are carried out by the OECD AI Policy Observatory\footnote{\url{https://www.oecd.ai/}}, and by the European Commission's AI Watch\footnote{\url{https://knowledge4policy.ec.europa.eu/ai-watch_en}} at the European level. However, academics working in responsible AI are most often locked in an information bubble of articles on XAI that are discussed at prestigious conferences, journals, and on preprint servers such as arXiv\footnote{\url{https://arxiv.org/}}. This interaction is complicated by the different aims of the stakeholders developing AI solutions versus the stakeholders that use AI solutions, who typically do not understand their limitations.

We know that there is a gap between the expectations (enshrined in strategies and regulations) and the reality (presented in research papers) related to AI \citep{krafft_defining}. And due to the increasing number of documents, it is close to impossible to analyze this gap by manually analyzing all source documents.

\subsection{Our Contribution}
To address this problem, we need a standardized knowledge base that can be processed in an automated way. In this paper, we present the concept and implementation of such a framework. To achieve that, we build a set of tools for scrapping, filtering, and preprocessing relevant documents. Our system extract information from documents using Natural Language Processing (NLP). The proposed framework processes not only AI regulations, which have been developed relatively recently, but also guidelines, whitepapers, and academic articles. To study the dynamics of influence between academia and policymakers, we must detect interconnections between papers and policy documents, both explicit (citations, references) and implicit (similarities in approach to concepts, same authors affiliations). The tools developed in this study allow following the process of institutionalizing ideas on how technologies associated with artificial intelligence should be regulated. 

The described system is, to our knowledge, the first such solution that combines research papers, strategies, and regulations with rich annotations. In the second part of the paper, we showcase a set of analyses that such a system can perform. However, this is by no means an exhaustive list of use cases. In this work, we focused on XAI papers, but the proposed method could be used more broadly to analyze any subfield of AI. We believe that this work creates the foundation for future analyses of cross-dependencies between strategies, articles, and regulations.

%The paper attempts to show how these ideas were first expressed in academic papers, then they took the form of whitepapers and guidelines, only recently to be transformed into proper regulations. The study captures how often conflicting ideas have been spreading at the academy, government institutions, and private enterprises.
\subsection{Relation to social sciences}
Studying the developments of regulations regarding AI by using automated AI systems is interesting, not only because of practical reasons. This framework not only allows discovering the directions in which regulations develop. Additionally, it could be used to study a topic that has always been important for social science -- the relation between humans and technology. At least from the research conducted by \cite{Ogburn}, scholars are interested in studying how and how quickly the culture embraces new technology. How much time do we need to develop norms and rules telling us how we should understand the new technology, how we should use it, and how we should be punished for not following the rules. Therefore no matter how extraordinary for us technologies associated with AI are, the type of problems they pose on the general level is not new. What is new is the possibility to use new technology to study its cultural embracing.

Social science offers a variety of ways of studying relations between technology and culture -- the system presented in this paper is built around a take on this issue coming from political science, more specifically, the studies on public policies. On the one hand, our system can extract the significant characteristics of AI policies. On the other hand, it can grasp the dynamic behind the process of shaping these policies. It allows studying policy design~\citep{siddiki_2020} and policy process~\citep{Weible} at the same time. The system enables us to analyze which sets of AI experts influence policies towards AI and, at the same time, study the characteristics of these policies using Institutional Grammar (IG). There have been attempts on automated IG tagging using NLP~\citep{Rice2021}, but code is not available.

\subsection{Related work}
\label{sec:related_work}
Using text processing techniques to tackle political science issues has been in use for a while, but only recently has there been the adoption of modern NLP methods~\citep{glavas_computational_2019,hollibaugh_use_2019}. A recent example of automated policy texts analysis is~\cite{linder_text_2020}, where information extraction methods were used to mine similarities between public policies texts. 

There are also recent examples of combining network analysis and NLP in political science. Namely, ~\cite{zaytsev_entity_2019} used named entity recognition and the Chinese Whispers algorithm in a quantitative approach to identifying actors' coalitions in the influence of policymaking.

Studying dynamics of machine learning research has been catching interest lately~\citep{martinez-plumed_research_2021}; however, there has been no record of quantitatively analyzing such dynamics between academic papers and public policies. The topic of the influence of research over policy has been studied using traditional methodologies~\citep{newman_policy_2016}.
There were attempts at analyzing the relationship between policymakers and academia in the context of XAI \citep{krafft_defining, LANGER2021103473}; however, the methodology of such studies never included analyzing documents produced both by academia and policymakers

\subsection{IG as a novel approach to information extraction}
\label{sec:ig}
The Institutional Grammar (IG) was created to solve the discussion regarding one of the crucial issues in social science -- the nature of institution~\citep{ig_ostrom}, more specifically, how institutions regulate human behavior. However, it is now used mainly as an analytical tool in policy design studies. This type of research is focused on "the purposeful, functional, and normative qualities of public policies" \cite[p.~1]{siddiki_2020}, and it is especially concerned with "the content of policies and how this content is organized" \citep{siddiki_2020, SchneiderIngram}. Because the content of policies is expressed mainly through legal regulations, IG is mainly used to analyze legal regulations. The tool allows not only to develop research in political science, but it has also found applications in computer science ~\citep{Frantz2016}. IG's attractive feature seems to be its ability to transfer legal text into a computer-readable format. 

IG has been developing since its creation by \citeauthor{ig_ostrom}. Its most current version -- IG 2.0 -- is presented in a codebook written in cooperation between political and computer scientists ~\citep{ig_codebook}.

The basic unit of IG analysis is a statement. There are two types of statements: constitutive ones and regulative ones. Constitutive statements define crucial elements of a particular policy~\textit{For the purpose of this Regulation, ‘provider’ means a legal person that develops an AI system} where regulative statements provide information on which activities are allowed, forbidden, or obligatory in a particular policy setting ~\textit{The European Data Protection Supervisor may impose administrative fines on Union institutions.} Each type of statement could be parsed using proper IG components (see Table~\ref{tab:ig_components}). In the case of our examples, they should be parsed as follows:~\textit{For the purpose of this Regulation(AC), ‘provider’(E) means (F) a legal person that develops an AI system (P)} and~\textit{The European Data Protection Supervisor (A) may (D) impose (I) administrative fines (B).}

\begin{table}
\begin{center}
    \caption{IG main components depending on statement type (regulative or constitutive) based on~\cite[pp.~10-11]{ig_codebook}.}
    \label{tab:ig_components}
    \begin{tabular}{ |p{2cm}|p{3cm}|p{2.2cm}|p{3cm}| }
    \hline
    Regulative statements & Description &  Constitutive statements  & Description\\
    \hline \hline
    Attribute (A)& The addressee of the statement. &  Constituted Entity (E)& The entity being defined.\\
    \hline
    Aim (I)& The action of addressee regulated by the statement. & Constitutive Function (F)& A verb used to define Constituted Entity.\\
    \hline
    Deontic (D)& An operator determining the level of discretion or constraint associated with Aim. & Modal (M)& An operator determining the level of necessity and possibility of defining Constituted Entity.\\
    \hline
    Object (B)& The receiver of the action described by Aim & Constituting Properties (P)& The entity against which Constituted Entity is defined.\\
    \hline
    Activation Condition (AC)& The setting to which the statements apply. & Activation Condition (AC)& The setting to which the statements apply.\\
    \hline
    Execution Constraint (EC)& Quality of action described by Aim & Execution Constraint (EC)& Quality of Constitutive Function.\\
    \hline
    \end{tabular}
\end{center}
\vspace{-0.7cm}
\end{table}

The scope of an IG implementation into research depends on its goals. Our study on AI regulations follows those analyses where only some IG components were identified in legal texts and examined~\citep{Heikkila2018}. 

\section{The architecture of the MAIR framework}\label{sec:sect2}
To automatically analyze AI regulations' dynamics, we must first gather policy documents and academic papers, enrich them with relevant meta-information, and find interconnections between them. 
 
\begin{figure}[h!]
    \centering
    \includegraphics[width=\textwidth]{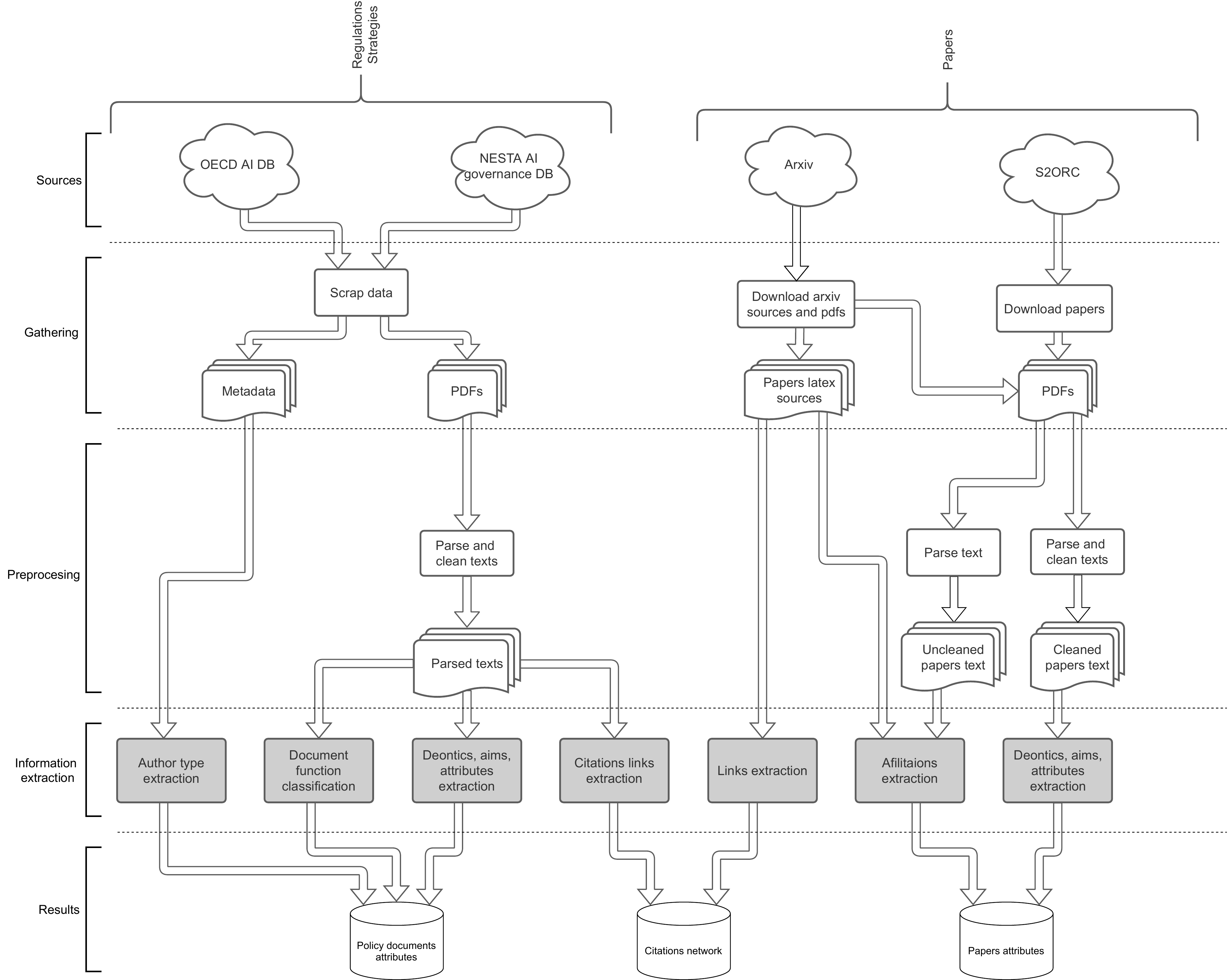}
    \caption{The process of acquiring and enriching documents for the MAIR database. The first level indicated by cloud icons identifies data sources, different for regulations and strategies and different for research papers. Subsequent components with a white background refer to the technical processing of the retrieved documents. The grey background indicates system elements that enrich documents with additional annotations, create additional meta-data or links between documents. The enriched documents are then stored in three databases describing different types of extracted data.} 
    \label{fig:system_pipline}
\end{figure}

The ideas described in sections \ref{sec:related_work} and \ref{sec:ig} inspired the development of the MAIR (Monitoring of AI Regulations, strategies, and research papers) framework. The architecture of this framework is shown in Fig.~\ref{fig:system_pipline}. The framework is fed with documents retrieved from four sources: OECD AI Policy Observatory\footnote{OECD AI Policy Observatory website: \url{https://oecd.ai/},  last download date: 19 Mar 2021.}, and NESTA AI Governance Database\footnote{NESTA AI Governance Database website: \url{https://www.nesta.org.uk/data-visualisation-and-interactive/ai-governance-database/}, last download date: 19 Mar 2021.} with policy documents and arXiv \citep{clement2019arxiv} as well as Semantic Scholar Research Corpus (S2ORC) \citep{lo-wang-2020-s2orc} for research papers. These documents are usually available as pdf files and are scrapped with Beautiful Soup\footnote{Beautiful Soup library is available on PyPi: \url{https://pypi.org/project/beautifulsoup4/}.}. 

System MAIR automatically detects some sections of texts, such as headers, and bibliography, to later extract citations and affiliations only from those parts of the text.
We extract and collect metadata, such as authors, source websites, and others, for later processing along with the content of documents. Then we run series of information extraction processes -- determine policy document function, extract deontic sentences along with Institutional Grammar attributes, determine authors and affiliations, find cross-citations between documents and other relevant data. All of those processes are described in detail in Sect.~\ref{sec.inf_extr}. All data gathering, processing, and extraction steps are managed by the DVC pipeline~\citep{ruslan-kuprieiev}, which allows for easy update of all results in case of new data available. 

The source code of the framework on the open GPL-3 license is available in the GitHub MAIR repository\footnote{GitHub MAIR repository:  \url{https://github.com/ModelOriented/MAIR}.}.

In the framework, we use two corpora of articles from arXiv:
\begin{itemize}
    \item \verb'arXiv.AI' that consists of all AI-related papers. These papers are identified based on the categories identified by authors (see Appendix \ref{arxiv-categories-appendix}). Due to its volume, this corpus contains only metadata. Today there are 164,105 documents in this corpus. This resource is used to identify papers referenced in policy documents in Sect. \ref{citation-network-construction}. \label{arxiv-AI}
    \item \verb'arXiv.XAI' that consists of a subset of the above related specifically to the domain of Explainable Artificial Intelligence and Interpretable Machine Learning, filtered by combinations of domain keywords (see Appendix \ref{xai-keywords}). It contains 742 papers with full texts along with metadata. Additionally, we extract a citation network by calling Semantic Scholar API.
 \label{arxiv-XAI}
\end{itemize}

\clearpage

%\subsection{Data sources}
%There are two primary document types we gathered - AI policy documents and research papers. The sources of policy documents are OECD AI Policy observatory (https://oecd.ai/) and NESTA AI Governance Database (https://www.nesta.org.uk/data-visualisation-and-interactive/ai-governance-database/).
%Documents are scrapped from those sources with Beautiful Soup ~\cite{richardson2007beautiful} using repetitive process. 
%As a source of academic papers we use arXiv \cite{clement2019arxiv}. We use two arXiv datasets in this work:
%\begin{itemize}
%    \item "arXiv AI" -- all AI-related papers based on a list of categories (Appendix \ref{arxiv-categories-appendix}). It contains only metadata rather than texts. The number of document in this dataset is 164,105. This resource is used to identify papers referenced in policy documents in \ref{citation-network-construction}. \label{arxiv-AI}
%    \item "arXiv XAI" -- a subset of the above related specifically to the domain of Explainable Artificial Intelligence, filtered by combinations of domain keywords (Appendix \ref{xai-keywords}). It contains 742 papers (texts along with metadata). \label{arxiv-XAI}
%\end{itemize}

 %Additionally, we gathered documents from semantic scholar research corpus \cite{lo-wang-2020-s2orc}, however, downside of this solution is that this data source is not constantly updated.

\section{Tools for knowledge extraction}
\label{sec.inf_extr}
This section describes NLP methods used to extract various characteristics from policy documents and academic papers. We extract document-wide qualities (document function, affiliations), links between documents, and a list of deontic sentences tagged with Institutional Grammar (IG) for every document. Scrapped metadata such as document issuing year are stored together with those extracted pieces of information and could be later used for the analysis of various aspects of dynamics. 
\subsection{Classification of document function}
Policy documents from Nesta and OECD fall into several categories based on their function. However, the classification provided by the authors is ambiguous and inconsistent between the two sources. We develop a more systematized classification system that we use as metadata in further analysis. Specifically, we define categories, perform manual annotation and train an NLP model for the automatic categorization.
For each document, we assign one of the following categories:
\begin{enumerate}
    \item \textbf{Diagnosis} -- reports, and other documents describing the current state of AI;
    \item \textbf{Principles} -- sets of ethical rules regarding AI;
    \item \textbf{Strategies} -- documents describing actions that should be taken towards AI;
    \item \textbf{Pre-regulations} -- proposals of legal regulations addressing AI;
    \item \textbf{Regulations} -- legal regulations addressing AI;
    \item \textbf{Body} -- documents establishing AI-related organizations.
\end{enumerate}
In the manual labeling process, we achieve 77\% agreement between two trained annotators and solve conflicts with a third annotator (the main author of this paper).
To classify documents, we use a few-shot learning model based on Task-Aware Representation of Sentences \citep{halder_task-aware_2020}, implemented in flairNLP \citep{akbik2019flair}.
We achieve 80.8\% accuracy on the holdout set. A detailed breakdown of accuracy per class is provided in Appendix \ref{sec:function_class_performance}.

\subsection{Extraction of Institutional Grammar (IG) tags}
In our solution, we focused on extracting 4 IG attributes from texts -- Attributes, Aims, Deontics, and Objects. Knowing that automatization of sentence tagging according to IG is not an easy task~\cite{Rice2021}, we simplified our approach to this analytical tool~\citep{Heikkila2018}. First of all, we tagged only these sentences which have modal verbs, because, through these type of sentences rights, obligations, and restrictions are usually expressed. Secondly, we treated all selected sentences as regulative statements. By doing this, we lost information if activities associated with deontic describe essential functions of entities to which statements are addressed or only potential actions they are capable of performing. Thirdly, we did not want to analyze very complex sentence structures, but we focused only on main sentences. Therefore our implementation of IG was tailored to the specific needs of our research.

The algorithm is based on dependency trees. The first step, after initial preprocessing, is splitting texts into sentences,, and parsing with spacy dependency parser \citep{spacy}. Then, we locate sentences with deontics from a closed list. The algorithm uses dependency relationships to locate finds verb (Aim), then subjects (Attribute) or passive subjects (Object). If there is no direct subject, we search the tree for clausal subject\footnote{For definition of clausal subject, see \url{https://universaldependencies.org/u/dep/csubj.html}} subsentence, and extract the subject of such subsentence. Then, any additional ObjectS are identified. We recursively repeat such a procedure for every verb that is in conjunction with the parent verb of deontic. In the end, we add every subject conjugated with any of the previously found subjects (same for objects). If any of found subjects is a pronoun, we perform the additional step of coreference resolution to find the entity to which the pronoun is referring. For this, we used Neuralcoref Spacy extension\footnote{Github repository with neuralcoref code: \url{https://github.com/huggingface/neuralcoref}}, which implements the method presented in  \citep{clark_deep_2016}.
Every deontic is then mapped onto one of the 3 categories: \textit{shall}, \textit{must} or \textit{can}, and every Object, Attribute, and Aim is lemmatized to simplify the further analysis. The details of the tagging algorithm are presented in the Appendix~\ref{sec:algorithm}.

\begin{figure}
\centering
\includegraphics[width=\textwidth]{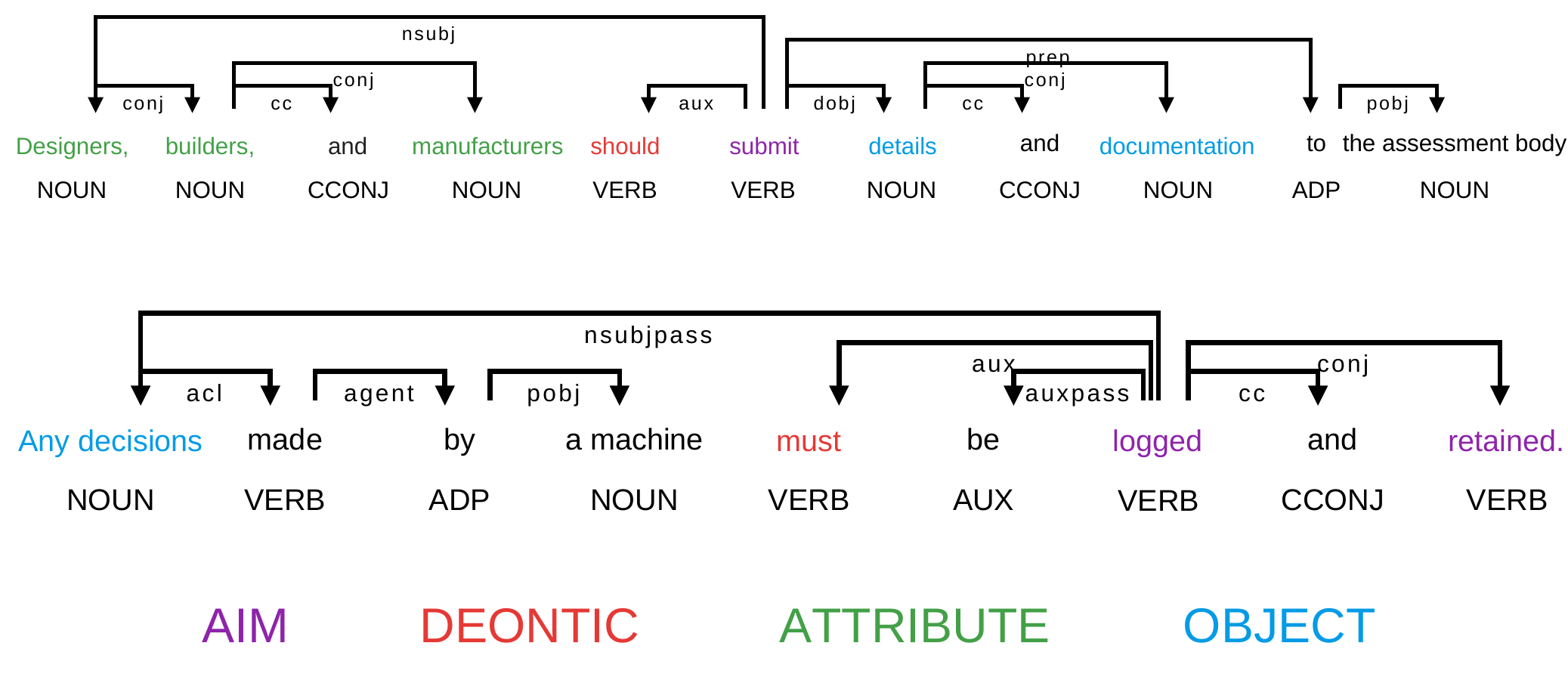}
% \includegraphics[width=\textwidth]{diagrams/complex_tree.pdf}
% \centering
% \includegraphics[width=\textwidth]{diagrams/passive_tree.pdf}
\label{fig:dependency}
\caption{Examples of dependency parsed deontic sentences.  Arrows represent dependencies, arrow labels are dependency relations types, and bottom words are part-of-speech tags -- all three are produced by the spacy parser and are used in the IG tagging algorithm. Colors are indicating IG tags assigned to words. In the first sentence, recognized Deontic (starting point for the algorithm) is "should" (which is translated in our nomenclature to "shall"). Our algorithm recognizes 3 Attributes ("designers", "builders", "manufacturers"), 1 Aim ("submit") and 2 Objects ("details", "documentation").
The second example is a passive sentence. The starting point for the algorithm is "must", there are two recognized Aims ("logged" and "retained") and 1 Object ("any decisions").}
\end{figure}

\subsection{Extraction of authors' affiliations }
One of the angles of our analysis is to discover players that actively impact the discourse and shape the AI regulations. To do this, we decided to extract the affiliations of the authors of papers. Here, we use an arXiv XAI dataset with sources of 742 papers (described in Sect. \ref{arxiv-XAI}).

Since arXiv collects metadata limited to the optional field and the submitter, the information is too sparse for any further application. For this reason, we extract it from the paper itself. Specifically, we choose to work on the LaTeX sources in a structured format. It is, however, not immediate to extract affiliations from this format, as multiple tags are enclosing this information. Additionally, there is no standard format for placing the affiliations in the paper, so they often mix with the author's names or the exact addresses. As a simplification, we do not intend to link specific authors with their institutions but instead, find a set of affiliations for the article.

Overall our pipeline consists of four steps:
\begin{enumerate}
    \item \textbf{Locate} the rough position of the affiliation in the text. We do it based on a list of identified LaTeX tags. Therefore we avoid extractions of institutions referred to in the text which are not authors' affiliations.
    \item \textbf{Extract} the names of the institutions.
    \item \textbf{Match} different names of one organization into one. For example, university name with/without the department
    \item \textbf{Classify} the organization as either academia or business.
\end{enumerate}

\begin{figure}
    \centering
    \includegraphics[width=\textwidth]{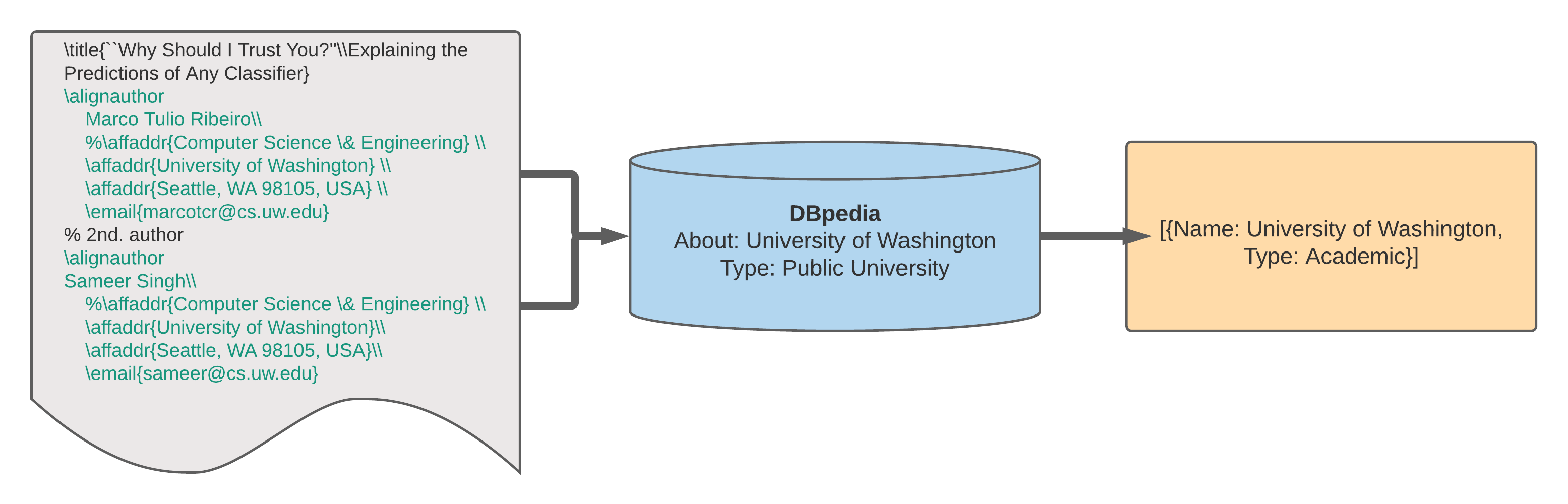}
    \caption{The process of affiliation extraction. In the Locate step, we use LaTeX tags to identify the potential location of the affiliations (left). Extract and Match stages performed with Named Entity Linking (NEL) result in matching with a corresponding DBpedia entry (middle). Classify step outputs type of affiliation based on DBpedia metadata (right).}
    \label{fig:affiliation_extraction}
\end{figure}

For various steps, we explored several options, including SpaCy \citep{spacy} Named Entity Recognition (NER) for extraction, utilizing the email domain as a proxy identifier of an institution for matching, and rule-based classification.
We then used Named Entity Linking (NEL)~\citep{6823700} for matching affiliations names with the external database. Specifically, we use the tool called Babelfy~\citep{moro-etal-2014-entity} which extracts entities and matches them against a DBpedia knowledge graph. Finally, we classify the affiliations based on their DBpedia entry tags.
A demonstration of the affiliation extractor run on Fig.~\ref{fig:affiliation_extraction}.

\subsection{Construction of citation network}
\label{citation-network-construction}
We constructed a citation network coupling research papers (arXiv) and policy documents (OECD, NESTA) focusing on references in policy documents pointing to academic papers. Our data consists of 196 policy documents and the \verb'arXiv.AI' dataset of 164,105 papers (described in detail in Sect.~\ref{arxiv-AI}).

\begin{figure}
    \centering
    \includegraphics[width=0.75\textwidth]{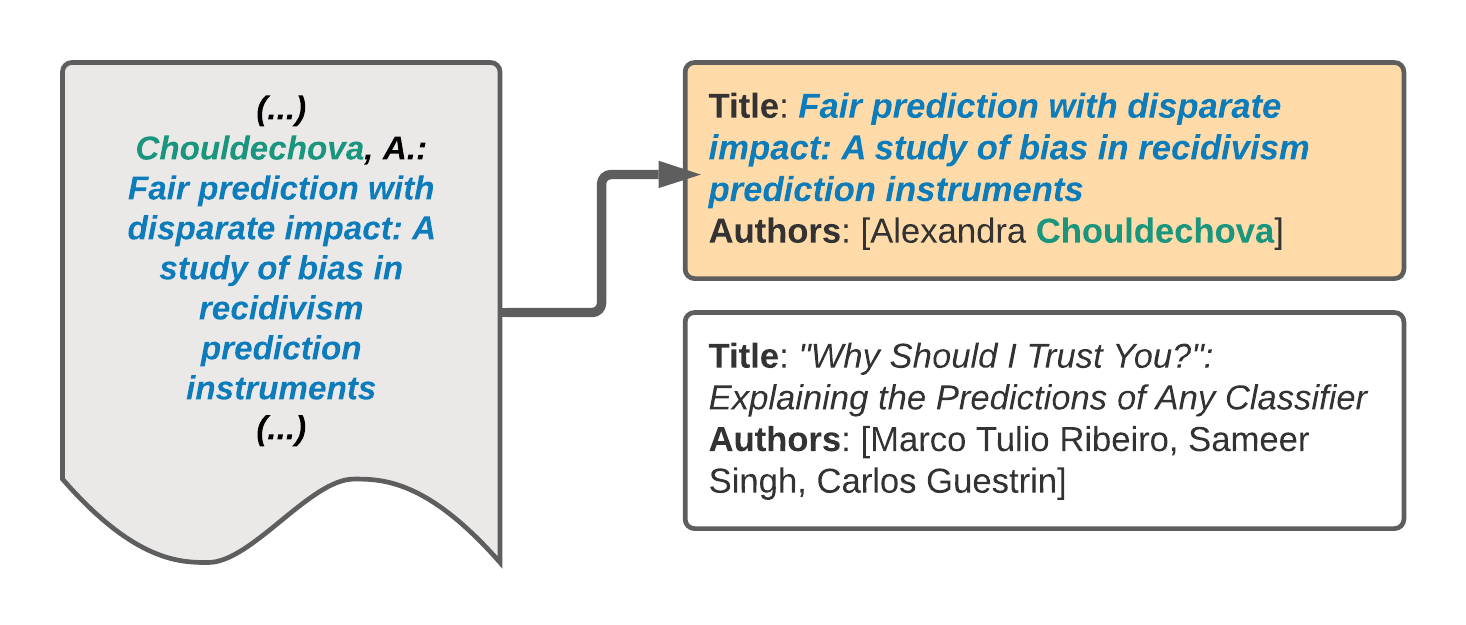}
    \caption{In the policy documents (left), we mine references pointing to research papers (right). A link is identified if there is a match in the metadata. In this example, the top of two papers is matched by a combination of title and author last name.}
    \label{fig:link_extractor}
\end{figure}

Policy documents do not contain any structured referencing format and are provided in the PDF format, which causes a lack of metadata about citations and prohibits us from using any of the existing tools assuming a consistent format of references.
We apply techniques from the field of Information Extraction to tackle issues of document linking~\citep{sil-etal-2012-linking} using metadata such as title, author names~\citep{shoaib2020author} in a free text \citep{essay73817}.
In this case, we pair each policy document with each paper and determine a match by either a paper's arXiv \verb'id' or a pair of (\verb'title', \verb'author') -- a demonstration of the link extraction method is shown in Fig.~\ref{fig:link_extractor}.

As a result, we obtain a bipartite graph of 202 links -- 37 policy documents citing 146 papers.

\section{Analysis of the MAIR corpus of documents}

Section \ref{sec:sect2} introduces the MAIR framework for collecting strategies, regulations, and research papers about XAI. Section \ref{sec.inf_extr} describes a set of techniques to enrich this corpus with additional meta-information extracted with advanced NLP tools. Such a corpus may serve many dedicated analyses related to the interdependencies between different stakeholders such as countries, IT companies, or academics. In this section, we present some example analyses of this corpus. It is not our aim to solve a specific research question but rather to show the versatility and usefulness of the developed resource. 

The first example focuses on the temporal analysis of citations between scientific articles and their cross-connections with policy documents; the second brings to the front inter-dependencies among different players -- in particular academia and industry. The third shows the use of deontic information to track differences in attitudes towards the human-AI relationship. Each of these examples describes an independent research problem. 

%To understand the interplay between research articles and policy documents we propose the analysis of the aforementioned citation network and deontics. 

%Such a network is a resource that aids understanding the influence of research world (papers, authors, institutions) on the AI policy documents and potentially vice versa. In order to bring to the front the affiliations detected in arXiv papers and track the inter-dependencies among them and OECD as well as NESTA policy documents, we shall consider here two different types of graph constructions:
%\begin{enumerate}
%    \item a network of citations among the selected XAI papers;
 %    \item a network created by the {\it bibliographic coupling} among the selected XAI papers. 
%\end{enumerate}

%On the other hand, the analysis of deontics allows to identify and compare main themes in scientific research and legal documents. It underlines the differences in attitudes towards selected topics and helps choosing objects for further inspection.

\subsection{XAI papers citation network}
\label{sec.arxiv_cit_an}
\begin{figure}[!ht]
    \centering
    \includegraphics[width=\textwidth]{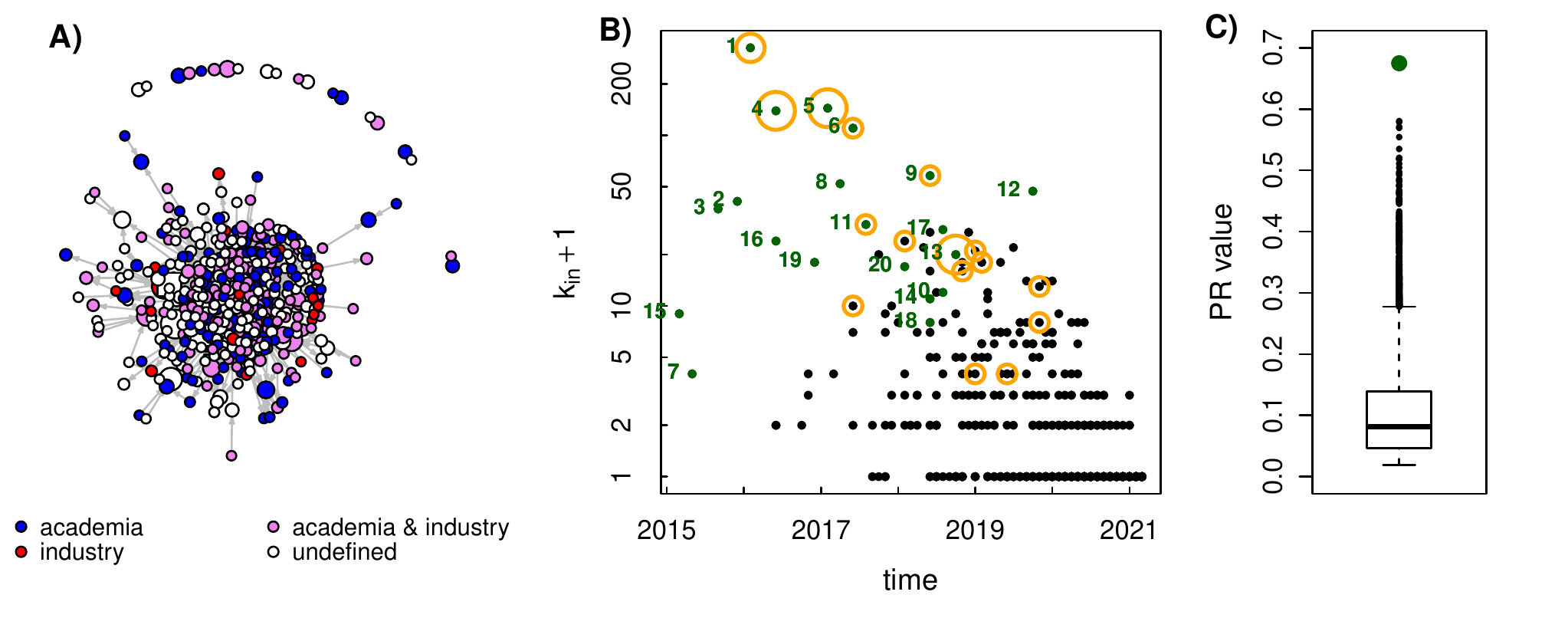}
    \caption{Analysis of XAI citation network $G_c$. A) XAI citation network with colour-coded nodes reflecting affiliation type (see legend); B) node in-degree $k_{in}$ vs publication time (as a log-linear plot is being used we plot $k_{in} + 1$ on the Y-axis); each black dot represent a single paper, 20 most influential papers according to Page Rank score are marked in green with numbers giving their rank, 16 papers cited by OECD and NESTA policy documents are marked with orange circles, their size being proportional to a number of such references; C) a box-plot of 10000 samples consisting of randomly (preserving time constraints imposed by citing policy documents) choosing nodes in $G_c$ and summing their Page Rank score, blue circle reflects he actual sum of Page Rank scores obtained for the cited papers.}
    \label{fig:netc}
\end{figure}

In the first case, we filter the \verb'arXiv.XAI' network of 742 arXiv papers (described in Sect. \ref{arxiv-XAI}) so that we take into account only nodes that cite (out connections) or are cited by (in-connections) at least one of the 742 papers. In result we restrict ourselves to a directed citation graph $G_c$ of $N_c=525$ nodes and $E_c=1919$ edges and no correlations among nodes' in- and out-degrees (Pearson's $r=-0.03$) consisting of one giant component with 505 nodes and 7 small clusters. The graph is shown in Fig.~\ref{fig:netc}A with color-coded nodes reflecting affiliation type. However, due to rather high density, the picture does not bring any specific insights. On the other hand, the analysis of the proportion of incoming and outgoing links reveals that the majority of connections goes to papers characterized affiliations identified both from academia as well as industry (see Table~\ref{tab:tabc}). There is also a significant difference in the profile of in-coming and outgoing links, e.g., although industry-affiliated papers have a very similar number of in- and out-connections (85 vs 87), it is almost three times less likely that an industry paper cites an academia one than an opposite situation to occur. Nonetheless, this picture might be dimmed by the significant number of not categorized papers.

\begin{table}[]
    \centering
    \caption{Breakdown of links in the citation graph $G_c$: rows give the number of outgoing connections while columns represent incoming ones. "Academia \& industry" means papers with affiliations from both academia and industry.}
    \label{tab:tabc}    \begin{tabular}{|c|cccc|c|}
    \hline
    out \textbackslash ~in & academia & academia \& industry & industry & none & $\Sigma$\\
    \hline\hline
  academia & 69 & 156 & 17 & 166 & 408\\
  both & 94 & 221 & 20 & 227 & 562\\
  academia \& industry & 6 & 32 & 3 & 46 & 87\\
  none & 118 & 314 & 45 & 385 & 862\\
  \hline
  $\Sigma$ & 287 & 723 & 85 & 824 & 1919\\
  \hline
  \end{tabular}
\end{table}

To find the most influential nodes in $G_c$ we have used \texttt{igraph} \texttt{R} package implementation of the Page Rank algorithm \citep{igraph} -- Fig.~\ref{fig:netc}B presents 20 top-ranked nodes marked on in-degree vs publication time plot. As expected, in general, Page Rank (which, for a given node, is the higher the more highly Page Rank nodes are pointing towards it)  promotes nodes representing earlier papers as they tend to accumulate citations reflected by the number of incoming links. In the following step, we have identified in $G_c$ 16 nodes that are being cited in total 23 times by 7 different OECD and NESTA policy documents -- they are marked with orange circles in Fig.~\ref{fig:netc}B, their size representing a number of obtained citations by different policy documents. 

As nearly half of the cited nodes are among the most influential ones, this allows for setting a hypothesis stating that policy documents tend to point to important scientific papers rather than selecting articles not fully recognized in the field. To test this hypothesis,  we define $PR$ as the sum of Page Rank score of 23 randomly selected nodes in $G_c$ keeping time constraints imposed by the publication date of policy documents (i.e., if a given policy document was published in 2020, we take into account only arXiv papers prior to that year). The results of 10000 repetitions are shown in Fig.~\ref{fig:netc}C in the form o box-plot as compared to the actual sum of Page Rank scores obtained for the cited papers (blue circle in Fig.~\ref{fig:netc}B), proving that the cited articles are, in fact, much more influential than a random set.

\begin{figure}
    \centering
    \includegraphics[width=\textwidth]{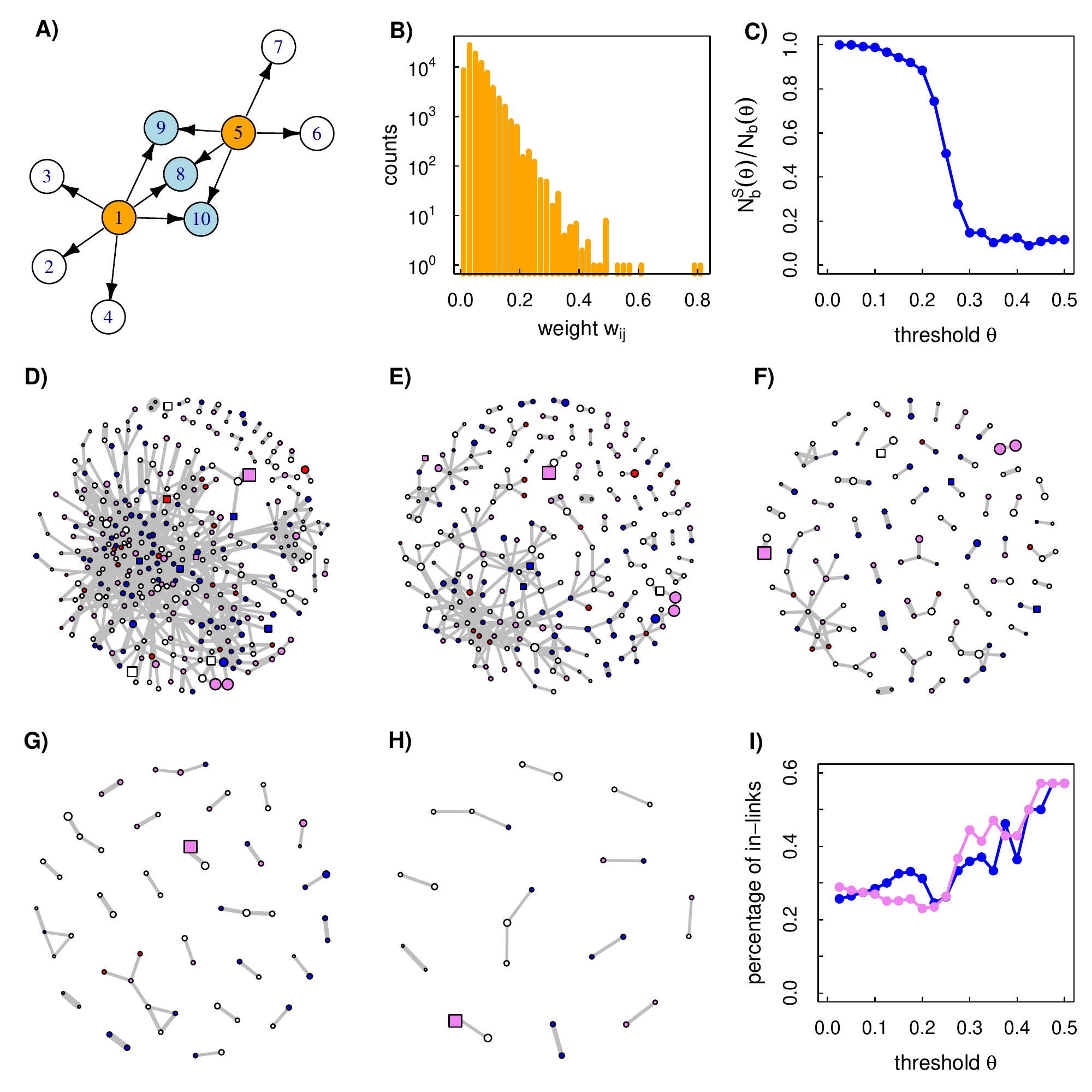}
    \caption{Aspects of XAI bibliographic coupling network. A) Explanation of the bibliographic coupling: two papers (here denoted as 1 and 5) cite, respectively, articles 2,3,4,8,9,10 and 7,6,8,9,10, only three papers out of 8 overlap, thus $w_{15}=3/8$; B) Histogram of the link weight $w_{ij}$ in $G_b$; C) Size of the giant component $N^S_b(\theta)$ normalized by the total number of nodes $N_b(\theta)$ for a given threshold value $\theta$; D)--H) Bibliographic coupling networks for different values of threshold $\theta$, respectively, 0.2, 0.25, 0.3, 0.35, 0.5. Node colors represent affiliation type as in Fig. \ref{fig:netc}A. The size of the node scales logarithmically with the Page Rank score obtained for the smaller set analysis, and the node shape informs if the paper has been cited by a document (rectangle) or not (circle); I) Percentage of homogeneous links academia--academia (blue) and academia \& industry--academia \& industry (violet) versus threshold $\theta$.}
    \label{fig:netb}
\end{figure}

\subsection{XAI bibliographic coupling network}

The graph described in the previous section represents actual links among arXiv papers. However, such a network is usually sparse and comes with a simple binary answer: either a paper is cited or not. To modify this approach we chose the so-called bibliographic coupling introduced by \cite{Kessler1963} which is simply the Jaccard index of out-neighborhoods $n_i^{out}$ and $n_j^{out}$ of two papers $i$ and $j$ \citep{Steinert2016}, i.e.,
\begin{equation}
    w_{ij} = \frac{|n_i^{out} \cap n_j^{out}|}{|n_i^{out} \cup n_j^{out}|}.
\end{equation}
The idea of bibliographic coupling is depicted in Fig.~\ref{fig:netb}A: in this way, we can take into account all information available in the references of each arXiv paper, unlike in the previous case. Additionally, we deal with a network where each link connecting nodes $i$ and $j$ is characterized with a weight $w_{ij} \in [0;1]$ reflecting similarity between two papers. The resulting bibliographic coupling undirected graph $G_b$ consists of $N_b=725$ nodes and $E_b=85598$ edges (in fact there should be $262450$ edges but we omit those carrying $w_{ij}=0$) with weight distribution roughly following an exponential function (see Fig. \ref{fig:netb}B). If follows that we are now able to use the concept of the {\it weight threshold} \citep[e.g.,][]{Chmiel2007} by keeping only such edges $w_{ij}$ whose weights fulfill the condition $w_{ij} \ge \theta$, where $\theta$ is a threshold parameter and $\theta \in [0,1]$. Such a procedure simply transforms the weighted graph $G_b$ into a set of unweighted networks $G_b(\theta)$ each constructed for a given parameter $\theta$ that are then subject to further analysis \citep{Sienkiewicz2018}. In particular, for some specific (critical) value of $\theta = \theta_c$ the network, initially percolated (i.e., constructed in such a way that is possible to arrive from any node $i$ to any other node $j$), breaks down into several small components. To track this phenomenon in a quantitative way for each $G_b(\theta)$ we calculate the size of its giant component $N^S_b(\theta)$ (i.e., the largest cluster in the network) and divide it by the relevant graph size $N_b(\theta)$. Figure~\ref{fig:netb}C allows localizing the breakdown at roughly $\theta_c \approx 0.25$, which can be visualized by a set of graphs in Fig.~\ref{fig:netb}D--H that not only reflect this process but also present other properties of the network such as affiliation (node color), importance (node size) or relation to policy documents (node shape). By increasing $\theta$ we bring to the front the strongest connections in the network (e.g., Fig.~\ref{fig:netb}H), which tend to be in the majority homogeneous as seen in Fig.~\ref{fig:netb}I where the share of academia and academia \& industry in-links are plotted against $\theta$. Contrary to that, homogeneous industry links are seldom observed and not likely to survive the introduction of high thresholds.  

%\subsubsection{Conclusions}

The citation network analysis presented in Sect. \ref{sec.arxiv_cit_an} suggests that key players in the extracted arXiv papers network are recognized as relevant from the policy documents perspective. On the other hand, when we turn to bibliographic coupling graphs, we can spot the persistence of homogeneous links among academia-like nodes that overtake the graph when the weakest connections are filtered out.
\subsection{Deontic analysis}
In this chapter, we present an example of a deontic analysis of documents from the MAIR database. We processed both legal documents and scientific papers so as to extract Attributes, Aims, Deontics, and Objects from individual sentences. 

\begin{figure}
    \centering
    \includegraphics[width=0.49\textwidth]{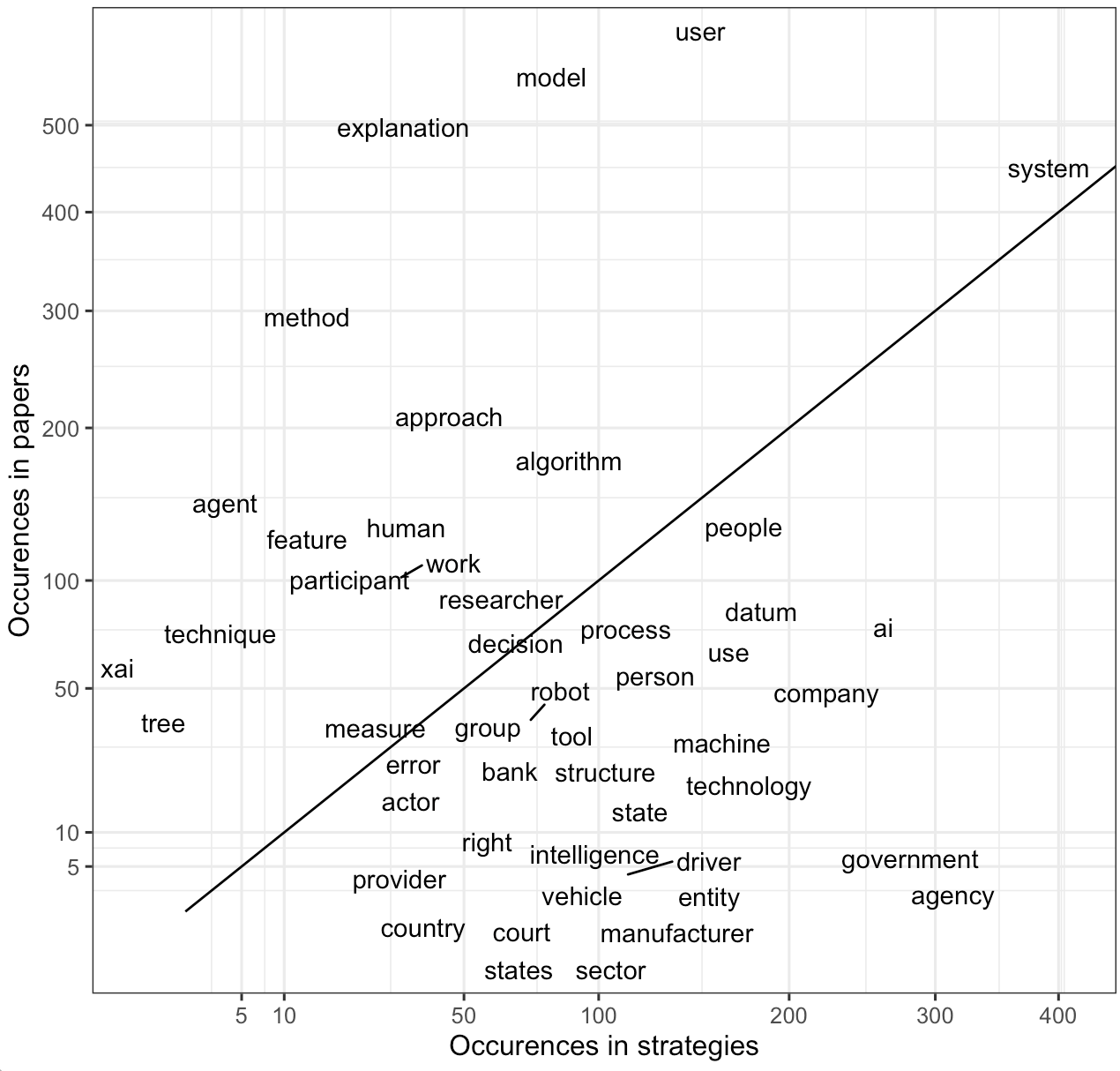}
    \includegraphics[width=0.49\textwidth]{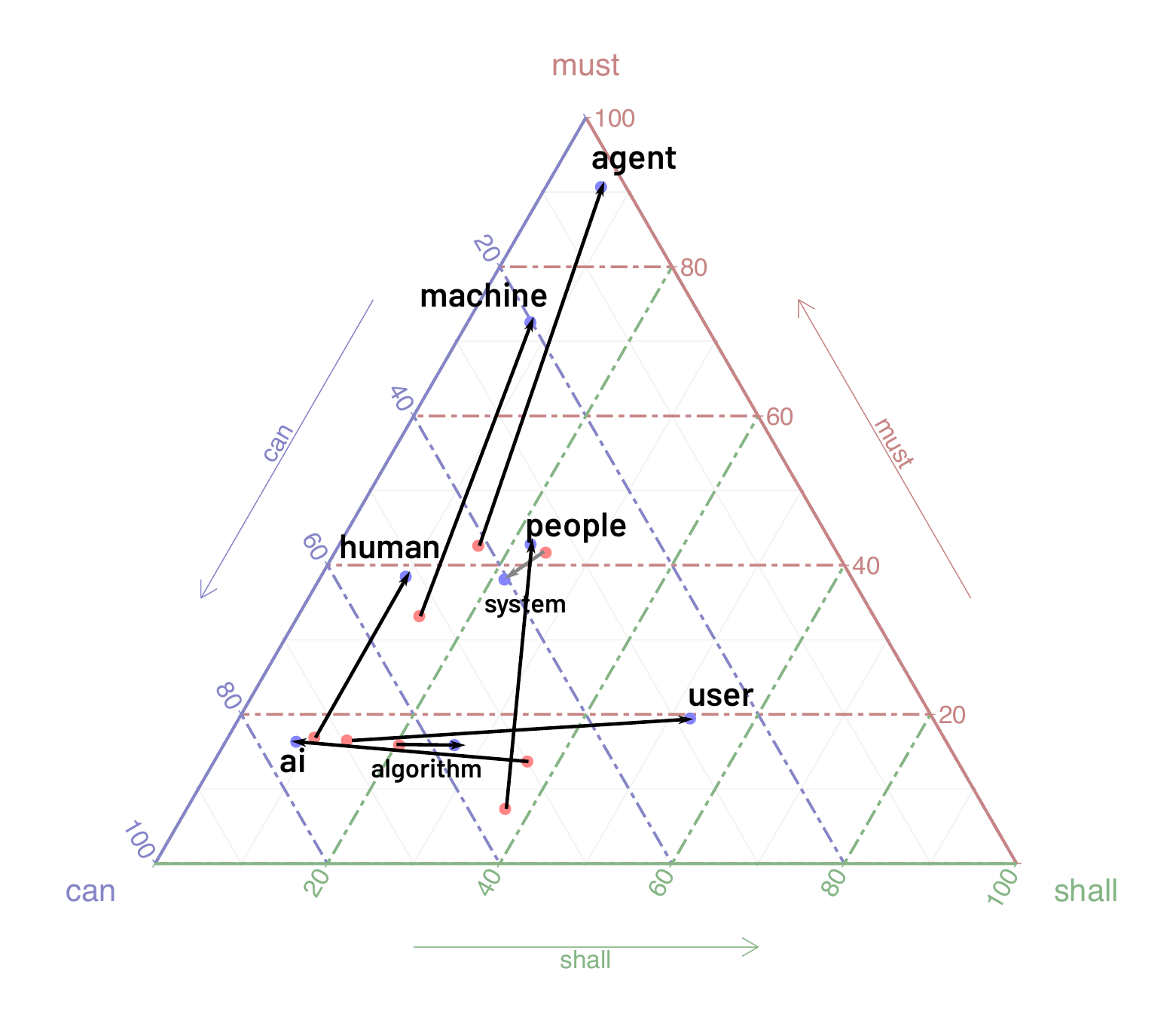}
\caption{Panel A describes the frequency of occurrence of each Object in research articles vs strategies. Only objects with more than 40 occurrences. Panel B presents for the selected eight objects the normalized context in which they are found in scientific articles (red dots) and strategies (blue dots). For the objects 'human', 'machine', 'agent', we can see a shift from 'must' in strategies to 'can' in scientific articles. For the object 'user', we have a shift along the dimension 'shall'.}

    \label{fig:deontic-panels1}
    \includegraphics[width=0.95\textwidth]{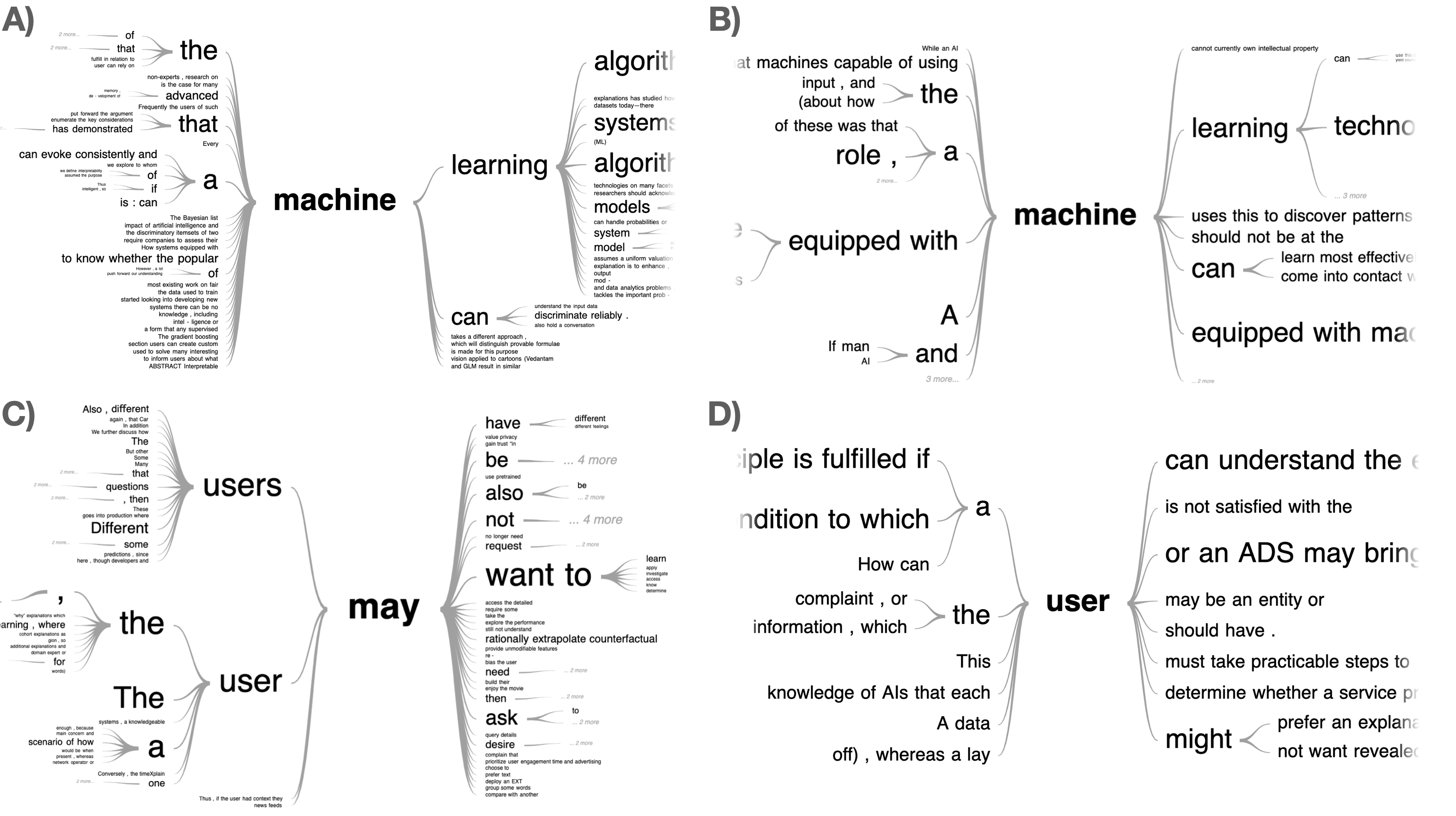}
\caption{Contexts of selected words in the analysed texts. Panels B and D show the context from scientific articles and panels A and C from strategies.}
    \label{fig:deontic-panels2}
\end{figure}

Panel A in Fig.~\ref{fig:deontic-panels1} shows how often the analysis of legal documents and academic papers identified a particular word as an object according to institutional grammar. Although the global size of both corpora of text was comparable, we can find objects which were much more frequently identified in the case of scientific publications (\verb'model', \verb'method', \verb'explanation', \verb'agent', \verb'feature') as well as those which are much more frequently encountered in legal documents (\verb'sector', \verb'government', \verb'agency'). A particularly interesting situation concerns the words \verb'driver' and \verb'vehicle', which appear very often in regulations and other legal documents, much less frequently in scientific publications. This may suggest that the topic of autonomous cars has a much stronger impact on the imagination of policymakers since many legal documents are devoted to it. For the XAI research community, it is not a foreground topic.

Based on the frequency of occurrence, we identified eight objects that underwent further deontic analysis (agent, machine, human, ai, people, algorithm, user, system). For each object, we determined whether it is accompanied by a term from the can / shall / must group (sentences in which negations occurred were few in number and were excluded from this analysis). The normalized frequency of each deontic in relation to the object was then presented in panel B of Fig.~\ref{fig:deontic-panels1}. Normalizations were carried out separately for scientific articles and separately for legal documents. The normalization was intended to remove the effect of the different frequencies of deontics in each group of texts. The ternary plots show the relative frequency of each deontic with a given object for both scientific articles and legal texts. Interestingly, in the case of legal documents, the word "AI" occurred more frequently near the deontic "can", the word agent or human near the deontic "must", and the word user near the deontic "shall". Such a shallow analysis allows for orientation in the area of global attitudes towards specific objects. At the same time, we see that the same objects occur in other contexts in the case of scientific papers. The object "user" definitely occurs more frequently in the context of the deontic "can", as do "machine" and "agent". We can see that scientific articles more often emphasize capabilities than strategies. At the same time, we observe an opposite trend for the word AI, which in the case of scientific papers more often occurs with the deontic "shall".

Shallow global analysis of objects and deontics suggests what kinds of objects are interesting to analyze. Having selected interesting phrases, we can use word trees to show the context in which certain phrases occur. Figure \ref{fig:deontic-panels2} shows word trees for selected phrases for research papers and legal documents. This type of interactive data mining allows for the analysis of well-defined questions. But to identify interesting questions, it is useful to use institutional grammar.

\clearpage

\section{Conclusions and Discussion}
The number of policy documents related to AI, like strategies and regulations, is growing rapidly. The number of research papers dedicated to interpretable and explainable AI is increasing at a higher rate. 
Literature related to the XAI field is divided into several polarised communities, ranging from advocates of solutions that explain any black-box model to researchers arguing that XAI should not be used for high-stake decisions.  A different perspective is presented by researchers from companies offering ML products from researchers using these solutions and bearing responsibility for errors in their work. The number and variety of these documents make it almost impossible to keep track of them continuously. Yet understanding the relationship between available methods and regulators' expectations is critical to implementing responsible AI. 

This paper introduces a novel framework for the automated analysis of documents related to trustworthy AI. This system integrates a set of state-of-the-art solutions from the fields of natural language processing (NLP), institutional grammar (IG), and network analysis (NA).
Each of these solutions is used to enrich raw text documents with relevant meta-information.
In this work, we have also shown the collection of focused analyses that can be performed for enriched documents, allowing us to use both from deontic information, author's affiliation, or graph of references information between documents.

As the interest in regulating XAI is increasing, it is essential to monitor how well reception and understanding of XAI by policymakers align with visions of XAI methods creators. In the future, such a system can be used to contribute to better cooperation between XAI researchers and AI policymakers. E.g. we will quickly assess if public policies are highly influenced by methods developed in papers written by specific opinion leaders.

%able to extract information and links from both policy %documents and academic papers. This system gathers documents, processes them with those tools, and aggregates the results for analysis. Our work showed that such framework can be used for novel analysis of interactions between academia and policymakers, using techniques from network analysis and institutional analysis. 

%Wersja pierwszego akapitu po poprawkach Bartka. Moje interwencje językowe mogą dotyczyć kwestii merytorycznych, więc wrzucam je w formie komentarza.

%This paper presents the first system that enables an in-depth analysis of relations between academic papers and policy documents regarding AI and XAI. The system gathers documents, processes them with tools we have created, and analyzes data extracted by these tools. Our work showed that such framework can be used for novel analysis of interactions between academia and policymakers, using techniques coming from network analysis and coming from institutional analysis – Institutional Grammer. 

%Additionally, methodology developed in this study, can be used to answer multiple questions regarding state and direction of AI institutionalisation and broader -- dynamics between the development of technologies and public policies.

%On the technical level, the system develops policy studies in two dimensions -- it allows to upscale the studies of policy documents by collecting all relevant research and policy papers regarding AI. It also allows performing these analyses almost in real-time. Our database of documents could be updated very easily.

\subsection{Limitations and future research}

Our system is now using a very simplified version of the Institutional Grammar tagger and shallow analysis of documents. In our future research, we would like to extend it by improving its ability for deep analysis of sentences with all their internal complexities. We also would like to be able to distinguish regulative statements from constitutive ones. 

Another limitation of our system is the relevance of policy documents -- we gathered them from databases that are manually updated. This limits our ability to draw conclusions from our analysis. To overcome these limitations, we should gather documents directly from relevant websites. What is more, to comprehensively analyze and understand the process of setting up regulations on XAI and AI, the system should also gather and process ethical guidelines on AI of private companies and even newspaper articles regarding XAI and AI. These documents are often cited in policy documents and influence the formulation of rules on new technologies.

Let us also mention that the analysis of both types of networks is directly affected by affiliations and references extraction methods. In effect, as can be seen in Table~\ref{tab:tabc}, affiliations of several nodes are labeled as ``none'', which introduces high uncertainty to the analysis of link homogeneity seen in Fig.~\ref{fig:netb}. Similarly, XAI papers' references are limited to arXiv papers only, which can influence both weight distribution and a relation among nodes in $G_b$. Future plans for the use of complex network analysis include identifying relation types among the papers based on the way they appear in the text \citep{Catalini2015} and examining different types of nodes' influence \citep{Lu2016} and citation measures \citep{Steinert2016}.       

\begin{acknowledgements}
Work on this project is financially supported by the NCN Sonata Bis-9 grant 2019/34/E/ST6/00052 \\
We are grateful to Anna Wróblewska for helpful discussions, and to Hubert Baniecki, Tomasz Stanisławek, and Krzysztof Kowalczyk for providing feedback on an early version of this paper.
\end{acknowledgements}

\bibliographystyle{spbasic}      % basic style, author-year citations
\bibliography{references}  %%% Remove comment to use the external .bib file (using bibtex).

\clearpage
\appendix
\section{Institutional Grammar tagging}
\label{sec:algorithm}
Below we present the pseudocode of the IG tagging algorithm. The implementation is available in our GitHub repository. 
\begin{algorithm}
\begin{algorithmic}
\caption{Extraction of IG tags from deontic sentence}

% \Procedure{GetSubjectsAndVerbs}{$ \var{deontic} $}
\State $\var{attributes} \gets \varnothing, \var{object} \gets \varnothing, \var{verbs} \gets \varnothing  $
\State $\var{verb} \gets \var{deontic}.parent$
\While{$\var{verb} \neq \varnothing $}
    \State $\var{attr} \gets \var{verb}, \var{verb} \gets \varnothing $
    \State $\var{verbs}.\func{append}(\var{attr})$
    \State $ \var{newSubj} \gets\{\var{c} : \var{c}\in \func{children}(\var{attr}), \func{relation}(\var{attr}, \var{c})=\func{nsubj}\}$
    \State $\var{newPassiveSubj}\gets\{c : c\in \func{children}(\var{attr}),  \func{relation}(\var{attr}, \var{c})=\func{nsubjpass}\}$
    \If {$\var{newSubj}=\varnothing \land \var{newPassiveSubj}=\varnothing$} 
    \Comment{Searching for clausal subject}
        \State $\var{attributes} \gets \{ \var{clausal}: \var{c}\in \func{children}(\var{attr}),\func{relation}(\var{attr}, \var{c})=\func{csubj}, \var{clausal} \in \func{children}(\var{c})\}$
    \EndIf
    \State $\var{attributes}\gets \var{attributes} \cup \var{newSubj}$
    \State $\var{objects} \gets \var{objects} \cup \var{newPassiveSubj} \cup \{\var{c} : \var{c}\in \func{children}(\var{attr}), \func{relation}(\var{attr}, \var{c})=\func{dobj}\} $
    \If{$\func{relation}(\var{attr}.parent, \var{attr})=\func{conj} \land \var{attr}.POS = \func{VERB}$}
        \State $\var{verb} \gets \var{attr}.parent$
    \EndIf
\EndWhile
\ForAll{$\var{subject} \in \var{attributes}$}
% \Comment{Adding all subjects that are in conjunction with any previous subject}
    \If{$\func{relation}(\var{attributes}.parent, \var{attributes})=\func{conj}$}
        \State $ \var{attributes}.\func{append}(\var{subject})$
    \EndIf
    \State $\var{attributes} \gets \var{attributes} \cup \{\var{s} : \var{s} \in \func{children}(\var{subject}), \func{relation}(\var{subject}, \var{s})=\func{conj}\}$
    \If{$\var{subject}.POS=PRONOUN$}
        \State $\var{subject} \gets \func{resolveCoreference(\var{subject})}$
    \EndIf
\EndFor
\ForAll{$\var{object} \in \var{objects}$}
    \State $\var{objects} \gets \var{objects} \cup \{\var{s} : \var{s} \in \func{children}(\var{object}), \func{relation}(\var{object}, \var{s})=\func{conj}\}$
\EndFor
\State \Return \var{attributes}, \var{objects}, \var{aims}
% \EndProcedure
\end{algorithmic}
\label{alg:ig_extraction}
\end{algorithm}
\clearpage

\section{Document function classification evaluation details}
\label{sec:function_class_performance}
\begin{figure}[H]
    \includegraphics[width=0.9\textwidth]{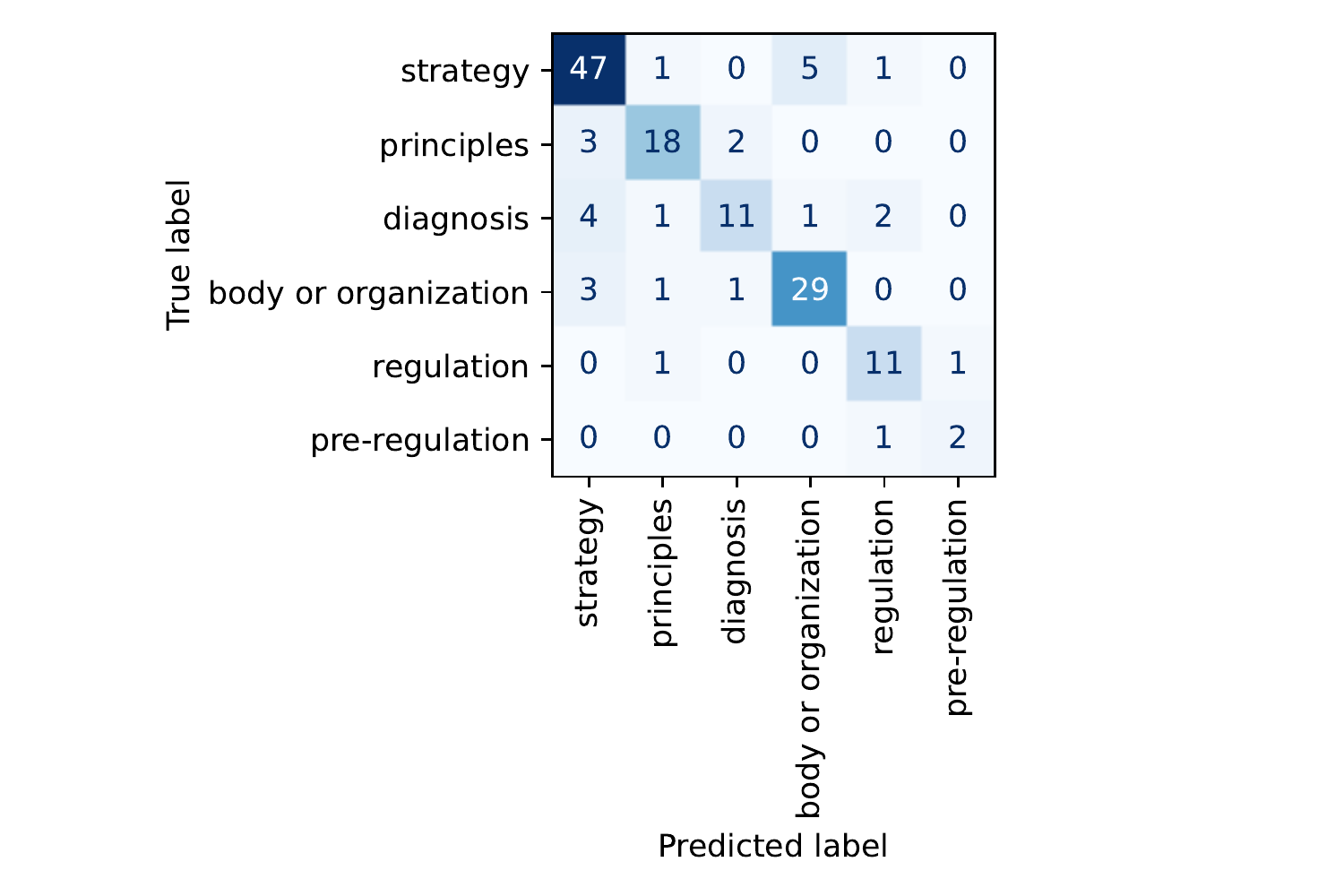}
    \caption{Confusion matrix of automatic classification of documents functions of our model on test set.}
    \label{fig:functions_confusion_matrix}
\end{figure}

The document classification model was trained on 596 and evaluated on 146 document titles. In Figure~\ref{fig:functions_confusion_matrix} we present an amount of misclassified items in the train set, divided by category.
\clearpage

\section{arXiv AI categories}
\label{arxiv-categories-appendix}
List of categories of arXiv papers used in the \verb'arXiv.AI' dataset. Note that some articles belong to multiple categories. Hence, the total is above the overall count. Access date: \emph{08/05/2021}.

\begin{table}[H]

\caption{List of articles categories along with corresponding number of articles in the \texttt{arXiv.AI} dataset.}
\begin{tabular}{ |c|c|c| } 
\hline
Category name & Category id & Number of articles \\
 \hline
Computer Vision and Pattern Recognition & cs.CV & 56,751 \\ 
 Machine Learning & stat.ML & 48,892 \\
 Artificial Intelligence & cs.AI & 33,214 \\ 
 Computation and Language & cs.CL & 26,610 \\ 
 Computers and Society & cs.CY & 10,218 \\
 Neural and Evolutionary Computing & cs.NE & 9,486 \\
 Computer Science and Game Theory & cs.GT & 7,458 \\
 \hline
\multicolumn{2}{|c|}{Total} & 164,105 \\
 \hline
\end{tabular}
\end{table}

\section{XAI keywords}
\label{xai-keywords}
List of keywords of arXiv papers used in the \texttt{arXiv.XAI} dataset. Matching is performed using arXiv API on titles, abstracts, and journal references. Note that some articles are matched with multiple keywords. Hence, the total is above the overall count. Access date: \emph{25/03/2021}.

\begin{table}[H]
    \caption{List of keywords along with corresponding number of articles in the \texttt{arXiv.XAI} dataset.}
    \begin{tabular}{|c|c|}
    \hline
    Keyword & Number of articles  \\
    \hline
    Interpretable Machine Learning & 458 \\
    Explainable Artificial Intelligence & 446 \\
    Fairness & 162 \\
    Transparency & 91 \\
    \hline
    Total & 742 \\
    \hline
    \end{tabular}
\end{table}

\end{document}